\documentclass[12pt]{article}
\usepackage{makeidx}
\usepackage{epsfig}
\usepackage{mathrsfs}
\usepackage{latexsym}
\usepackage{amsfonts}
\usepackage{amssymb}
\usepackage{amsmath}
\usepackage{amsthm}
\usepackage{appendix}
\usepackage{color}
\usepackage{xcolor}
\def\be{\begin{equation}}
\def\ee{\end{equation}}
\def\bea{\begin{eqnarray}}
\def\eea{\end{eqnarray}}
\begin{document}
\title{
{\bf D-brane description from nontrivial M2-branes}
}
\author{
\hspace*{-20pt} \small \bf M.P. Garcia del
Moral, C. Las Heras\footnote{These authors have contributed equally to this work.}\\
{\small Departamento de F\'\i sica, Universidad de Antofagasta}\\[-6pt]
{\small Antofagasta, Chile} \\
}
\maketitle

\begin{abstract}
We obtain the bosonic D-brane description of toroidally compactified non-trivial M2-branes with the unique property of having a purely discrete supersymmetric regularized spectrum with finite multiplicity. As a byproduct, we generalize the previous Hamiltonian formulation  to describe a M2-brane on a completely general constant quantized background $C_3$ denoted by us as CM2-brane. We show that under this condition, the theory is equivalent to a more restricted one, denoted as an M2-brane with $C_\pm$ fluxes, which has been shown to have good quantum behavior. As a result, the spectral properties of both sectors must be the same.We then obtain its bosonic D-brane description and discover new symmetries. We find that it contains a new symplectic gauge field in addition to the expected U(1) Dirac-Born-Infeld gauge symmetry and a nontrivial $U(1)$ associated with the presence of 2-form fluxes. Its bundle description takes on a new structure in the form of a twisted torus bundle. By turning off some of the fields, the D-brane description of other toroidally nontrivial M2-brane sectors can be obtained from this one. The possibility of reinterpreting these sectors in terms of Dp-brane bound states is discussed.  These D-brane descriptions constitute String theory sectors with a quantum consistent uplift to M-theory. 
\end{abstract}
\clearpage
\tableofcontents

\section{Introduction}
The Supermembrane theory  was  originally expected to describe the microscopic degrees of freedom of M-theory, however when formulated on 11D  Minkowski background it has a continuous spectrum from $[0,+\infty)$
\cite{deWit6,deWit2}. This behavior does not change just by circle compactifications \cite{deWit3}. The continuity of the spectrum  represents an obstruction to interpreting the M2-brane as a fundamental object. Indeed, it led to the formulation of the matrix theory conjecture \cite{Susskind}  where the supermembrane was interpreted as a second quantized theory. In \cite{Restuccia}, a new topological condition was found associated to the presence of a non trivial $U(1)$ bundle on the worldvolume of the supermembrane induced by an irreducible wrapping. In {\cite{Ovalle3}}, its Hamiltonian formulation was found. The spectrum of the regularized supersymmetric theory was rigorously shown to be purely discrete, with finite multiplicity \cite{Boulton} in distinction to the general case. This sector was denoted as Supermembrane (i.e. M2-brane) with central charges, since the topological term associated to an irreducible wrapping induces the presence of a non-vanishing central charge in the supersymmetric algebra. It has been formulated  for different target spaces \cite{Bellorin,mpgm14,mpgm15}. 

Other sectors of the M2-brane with good quantum properties have also  been identified: the M2-brane on a pp-wave background \cite{Sugiyama,Dasgupta} whose regularization is described by the BMN matrix model \cite{Maldacena3} and its spectral properties  proved in \cite{mpgm11}, as well as a toroidally compactified M2-brane in the presence of a quantized constant three-form that induces worldvolume  2-form fluxes \cite{mpgm6}. Recently, a new sector associated to the compactification of the M2-brane on a particular 10D spacetime with punctures has also been obtained. It can describe other type of non-trivial M2-branes \cite{mpgm16} once a suitable regularization is provided. The regularization of these non trivial M2-brane sectors satisfy the sufficiency condition for discreteness found in \cite{Boulton}. These M2-brane sectors describe \textit{part} of the fundamental M-theory degrees of freedom and therefore they may represent a restriction on the String sectors with a quantum consistent uplift to M-theory. The need to obtain other non-trivial sectors of the M2-brane theory formulated on more general backgrounds becomes increasingly clear, but also its String counterparts. 
 
 In this paper we concentrate on the String description of the nontrivial  M2-brane bosonic sector associated with the presence of a central charge condition or subject to a quantized constant three-form background that induces a 2-form flux condition on its worldvolume. The relation between the M2-branes compactified on target spaces with isometries and the D2-branes under scalar/vector dualization is well-known \cite{Ovalle3,Duff5,Townsend4,Schmidhuber,Polchinski3,Manvelyan1,Duff6, Bergshoeff7,Ovalle1,Leigh}. In this sense, sometimes these two theories have also been referred as duals. We will also use the world 'duality' in the aforementioned sense when we will refer to it in sections 5 and 6.  Some of the distinctive properties of these nontrivial M2-brane sectors will be inherited by their String duals, as we will see. In particular, the M2-brane with central charges and  the M2-brane with $C_{\pm}$ 2-form fluxes have a nontrivial U(1)  monopole connection and a symplectic gauge field which can be properly combined to produce  a topologically nontrivial dynamical gauge field as recently proved in \cite{mpgm10}. This extra field is not present in an ordinary supermembrane theory and it is therefore natural to expect that new fields may arise on the associated D-brane theory. Because this extra field does not exist in a standard supermembrane theory, it is natural to expect new fields to emerge on the associated D-brane theory. When Dp-branes contain worldvolume fluxes they can be expressed in terms of Dp-brane bound states  \cite{Maldacena4,Dasgupta2,Roy}. We will discuss this point  briefly at the end of the paper.\newline
 
The paper is organized as follows: In section 2, we review recent results about the main properties of two nontrivial M2-brane sectors: the M2-brane with central charges and the M2-brane with $C_{\pm}$ fluxes. We discuss their equivalence relation, which can be interpreted as a duality connecting those two M2-brane sectors, a priori inequivalents. In section 3, we obtain the M2-brane with a $C_{\pm}$  flux Hamiltonian where the transverse components of the supergravity three-form become explicit. We denote it as CM2-brane and characterize its properties. In section 4, we obtain its D-brane description. It contains RR and NSNS quantized charges that generate the presence of 2-form fluxes and a new dynamical field. In section 5, we discuss the physical properties of the new symmetries and their global description. In section 6, we show that it is possible to obtain the same result by departing from a D2-brane in $10D$ toroidally compactified when some specific fluxes are turned on and they generate an extra constraint that modifies the original Hamiltonian. We discuss its differences with respect to a D2-brane with RR and NSNS background fields with generic fluxes. In section 7, we present a brief discussion and our conclusions.

\section{Toroidally compactified Nontrivial M2-branes}
In this section, we will briefly review former results concerning  the local \cite{Ovalle3,Ovalle1,Ovalle2} and global aspects of nontrivial M2-branes associated with the presence of central charges \cite{mpgm7}, those of the M2-brane formulated in the presence of 2-form fluxes $C_{\pm}$ \cite{mpgm6,mpgm10} and their relationship. The Light Cone Gauge (LCG) $D=11$ bosonic Hamiltonian formulation of a M2-brane on a general background was found in \cite{deWit}. Its supersymmetric extension on a flat superspace coupled with a non vanishing constant supergravity three-form is given by  \cite{mpgm6}
\begin{eqnarray}\label{HCM2}
\mathcal{H}&=&\left[\frac{1}{(P_--C_-)}\left(\frac{1}{2}(P_a-C_a)^2+\frac{1}{4}(\epsilon^{uv}\partial_u X^a \partial_v X^b)^2\right) - \bar{\theta}\Gamma^-\Gamma_a \left\lbrace X^a,\theta \right\rbrace \right. \nonumber  \\
&-& \left. C_{+-}- C_+ \right]
\end{eqnarray}
subject to 
\begin{eqnarray}
P_a\partial_u X^a + P_- \partial_u X^- + \bar{S}\partial_u \theta &\approx& 0\\
S - (P_--C_-)\Gamma^- \theta &\approx& 0
\end{eqnarray}
where  $X^a,X^-$ denote the embedding maps from the worldvolume $\Sigma$, assumed to be a Riemann surface of genus one  onto the target space. The indices $a,b=1,\dots,9$ denote the target space transverse components and $u,v=1,2$ the spatial directions of  $\Sigma$. $\theta$ represents a Majorana spinor of 32 components that acts as a scalar on the worldvolume and $\Gamma$ are the Gamma matrices in 11D. The canonical conjugate momenta related to $X_a$, $X^-$, $\theta$, are given by $P_a$, $P_-$ and $S$, respectively. 
The LCG supergravity three-form components \cite{deWit} correspond to 
\begin{equation}\label{CaLCG}
\small
\begin{aligned}
& C_a  =  -\varepsilon^{uv}\partial_uX^- \partial_vX^b C_{-ab} +\frac{1}{2}\varepsilon^{uv}\partial_uX^b \partial_vX^c C_{abc} \,  \\
& C_{\pm}  =  \frac{1}{2}\varepsilon^{uv}\partial_uX^a \partial_vX^b C_{\pm ab} \,, \qquad C_{+-}  =  \varepsilon^{uv}\partial_uX^- \partial_vX^a C_{+-a} \,
\end{aligned}
\end{equation}
The gauge invariance of the three-form allows us to fix $C_{+-a}=0$ and $C_{-ab}=0$. In \cite{mpgm6} the authors restrict themselves to considering backgrounds with  $C_{\pm ab}$ and $C_{abc}$ the only nonzero and constant components. The Hamiltonian and the constraints contain nonphysical degrees of freedom associated to $X^-$ that must be eliminated \cite{deWit}. A way to solve this problem without introducing non-localities was proposed in \cite{mpgm6}, eliminating this dependence through the following canonical transformations of the phase space variables.
\begin{eqnarray}\label{TransCanonica}
\widehat{P}_a &=& P_a - C_a \quad , \quad
\widehat{P}_- = P_- - C_-
\end{eqnarray}
Indeed, it can be seen that these transformations preserve the kinetic terms and all the Poisson brackets. Furthermore, it can be set $\widehat{P}_-=P^0_-\sqrt{W}$, with $\sqrt{W}$ a regular density on the worldvolume $\Sigma$ corresponding to the determinant of the spatial part of the worldvolume metric. If the target space is now toroidally compactified to $M_9^{LCG}\times T^2$ the embedding maps $X^a (\sigma^1,\sigma^2, \tau)$ become decomposed in terms of the maps $X^m$ with $m=3,\dots,9$ from the base $\Sigma$ to $M_9^{LCG}$ and the maps $X^r$ with $r=1,2$ from $\Sigma$ to $T^2$.
The winding condition on the compact sector,
\begin{eqnarray}\label{M}
\oint_{C_{s}} dX^r = M_s^r 
\end{eqnarray}

such that $m_s^r$ are the elements of the winding matrix. In complex coordinates  $M^1_s+iM^2_s = 2\pi R(l_s+m_s\tau)$ with $R$, $\tau$ the moduli of the $T^2$ and $l_s,m_s$ the winding numbers, allows us to define $dX_{h}^r=M^r_sd\widehat{X}^s$ in terms of the orthonormalized harmonic basis  $d\widehat{X}^r$,  $$\oint_{C_s}d\widehat{X}^r=\delta_s^r$$ with $C_s$ the homological basis if the torus $T^2$. Since the components of the three-form are constants, they can always expressed as $C_{\pm rs}=c_{\pm} \epsilon_{rs}$ with $c_\pm \in \mathbb{Z}/\{0\}$.  It can be defined $\displaystyle \widetilde{F}_{\pm}=\frac{1}{2}C_{\pm rs}M^r_pM^s_q d\widetilde{X}^p\wedge d\widetilde{X}^q$  with $\widetilde{X}^p$, $p=1,2$, the $T^2$ coordinates. When a quantization condition is imposed on the three-form components $C_{\pm rs}$, then, the 2-form flux condition is defined \cite{mpgm6},
  \begin{eqnarray}\label{fluxpullback}
 \int_{T^2}C_{\pm}= \int_{T^2}\widetilde{F}_{\pm} =  k_{\pm}\in \mathbb{Z}, \quad k_{\pm}\ne 0.
  \end{eqnarray}
 The target space flux condition indices a worldvolume flux condition through its pullback. The minimal embedding maps $\widehat{X}^r$ are identified with the $T^2$ torus coordinates $\widetilde{X}^s$, and the wordvolume and target-space fluxes are in one-to-one correspondence  \cite{mpgm10}, 
   \begin{eqnarray}\label{fluxpullback}
 \int_{T^2}\widetilde{F}_{\pm} =  c_{\pm}\int_\Sigma \widehat{F}, 
  \end{eqnarray}
with $\displaystyle \widehat{F}=\frac{1}{2}\epsilon_{rs}dX_h^r\wedge dX_h^s$ defined in terms of the harmonic one-forms.The flux units are $k_{\pm}=nc_{\pm}$ with $n$ first Chern class associated to $\widehat{F}$. 
Once the flux condition is imposed and applying the Hodge decomposition the closed one-forms can be decomposed as,
$dX^r=M^r_sd\widehat{X}^s+d\mathcal{A}^s$ where $dA^s$ are components of an exact one-form that transform as a symplectic connection under symplectomorphisms transformations. 
The flux condition acts as a new constraint on the Hamiltonian that changes it. Then, the Hamiltonian of the M2-brane on a $C_{\pm}$ flux background becomes
\begin{eqnarray}
\label{hamiltoniancpm}
H^{\tiny{C_{\pm}}}&=&\int_\Sigma d^2\sigma\sqrt{W}\left[\frac{1}{2}\Big(\frac{\widehat{P}_m}{\sqrt{W}}\Big)^2+\frac{1}{2}\Big(\frac{\widehat{P}_r}{\sqrt{W}}\Big)^2 + \frac{1}{4}\left\{X^m,X^n\right\}^2 + \frac{1}{2}(\mathcal{D}_rX^m)^2 \right. \nonumber  \\
&+& \left. \frac{1}{4}(\mathcal{F}^{rs})^2+\frac{1}{4}(\widehat{F}^{rs})^2 -\bar{\theta}\Gamma_-\Gamma_r\mathcal{D}_r\theta-\bar{\theta}\Gamma_-\Gamma_m\left\{X^m,\theta\right\} +C_+\right]\ 
\end{eqnarray} 
where
\begin{eqnarray}
\left\lbrace A, B \right\rbrace = \frac{\epsilon^{uv}}{\sqrt{W}}\partial_u A \partial_v B
\end{eqnarray}
is the Lie Bracket as defined in \cite{deWit5} and $\widehat{F}^{rs} = \left\lbrace X_h^r,X_h^s \right\rbrace = \frac{1}{2}\epsilon^{rs}\frac{\epsilon^{uv}}{\sqrt{W}}\widehat{F}_{uv}$  with $\widehat{F}_{uv}$ the components of the worldvolume two-form $\widehat{F}$.
%
%$\widehat{F}^{rs}=\frac{1}{2}\epsilon^{rs}\frac{\epsilon^{uv}}{\sqrt{W}}\widehat{F}_{uv}$
%

 The action of the worldvolume flux constraint induces the appearance of a monopole contribution in the Hamiltonian and a dynamical symplectic gauge field $\mathcal{A}^r$ \cite{Ovalle3}. The degrees of freedom of the theory are $X^m,\mathcal{A}^r,\theta$. It also contains a  symplectic covariant derivative and a symplectic curvature that were both formerly identified in \cite{Ovalle1} given by 
\begin{eqnarray}\label{simplectica}
   \mathcal{D}_rX^m =D_rX^m+\left\{ \mathcal{A}_r,X^m\right\}, \qquad
   \mathcal{F}_{rs}= D_r\mathcal{A}_s-D_s\mathcal{A}_r+\left\{ \mathcal{A}_r,\mathcal{A}_s\right\}
\end{eqnarray}
with $D_r$ a covariant derivative such that $D_1+iD_2=2\pi R (l_s+m_s\tau)\Theta^{s}_r\frac{\epsilon^{uv}}{\sqrt{W}} \partial_u \widehat{X}^r\partial_v$. it is defined in terms of the torus moduli $(R,\tau)$, the windings $l_s,m_s$ and $\Theta$ is  a matrix associated with the monodromy induced on the base manifold which is contained in the conjugacy classes of $SL(2,Z)$ \cite{mpgm17}. Its presence is due to the invariance of the theory under the full symplectomorphisms group, in particular with those that are not connected with the identity on $\Sigma$ that changes the homology basis on $\Sigma$ and the corresponding basis of one-forms $d\widehat{X}$ \cite{mpgm10}. 

This Hamiltonian is subject to the local and global constraints associated to the Area Preserving Diffeomorphisms (APD) as a residual symmetry,
\begin{eqnarray}
\small \left\{ (\sqrt{W})^{-1}\widehat{P_m} , X^m\right\} + \mathcal{D}_r\left( (\sqrt{W})^{-1}\widehat{P}_r\right)+\left\lbrace (\sqrt{W})^{-1}\bar{S},\theta \right\rbrace  &\approx& 0
\label{APDconstraints1}\\
 \oint_{C_S}\left[\frac{\widehat{P}_m dX^m}{\sqrt{W}} + \frac{\widehat{P}_r dX^r}{\sqrt{W}} + \frac{\bar{S} d\theta}{\sqrt{W}}\right] &\approx& 0 \label{APDconstraints2},
\end{eqnarray}
The worldvolume flux condition acts as a new constraint on the Hamiltonian. 
%The imposed flux condition is equivalent to the existence of a $U(1)$ principle bundle over $T^2$ whose pullback generates a flux $\widehat{F}$  associated to the presence of monopoles on its worldvolume \cite{Restuccia2}.
 The regularized supersymmetric Hamiltonian \cite{mpgm6}  satisfies the sufficiency criteria for discreteness found in \cite{Boulton}. The theory is $N=1$ since it preserves $1/2$ of the supersymmetry \cite{mpgm6}. 

 \subsection{The M2-brane with $C_{\pm}$ fluxes dual to the M2-brane with central charge} 
  The so-called M2-brane with central charges \cite{Ovalle1} corresponds to an irreducibly wrapped toroidal M2-brane that contains an extra topological constraint \cite{Restuccia}  associated to an irreducible wrapping \begin{equation}\label{central charge}
   \int_{\Sigma}dX^r\wedge dX^s =\epsilon^{rs}nA_{T^2}
  \end{equation}
  with  the integer $n=det(\mathbb{W})\ne 0$ defined in terms of $\mathbb{W}$ the wrapping matrix and $A_{T^2}$ the 2-torus area. As it was shown in \cite{Restuccia2}, the irreducible wrapping condition represents a generalization of the Dirac monopole condition over Riemann surfaces of arbitrary genus $g\ge 1$. Classically, the dynamics of its associated Hamiltonian does not contain string-like spikes with zero cost energy \cite{mpgm}, which would render at quantum level, the supersymmetric spectrum of the theory continuous. The discreteness of its supersymmetric spectrum was rigorously proved in \cite{Boulton} and in \cite{mpgm11}. The central charge (CC) condition  (\ref{central charge}) corresponds to the quantization condition of $\widehat{F}$ and hence establishes a one-to -one relation with the quantization condition ($C_{\pm}$ over the 2-torus target space. In fact, $dX^r\wedge dX^s=\epsilon^{rs}\widehat{F}$ with $\widehat{F}$ defined in the previous section in such way that satisfies (\ref{central charge}).
  
  As shown in \cite{mpgm6} it establishes a relation between the two associated Hamiltonian densities as follows:
  \begin{eqnarray}
\mathcal{H}^{C_{\pm}} = \mathcal{H}^{CC} + C_+
\end{eqnarray}
  For $C_+=0$, the Hamiltonian with central charges and the one with $C_-\ne 0$ fluxes, exactly coincide. Hence the discreteness property of the former automatically implies the discreteness of the second one. For the case of $C_+\ne 0$, since it is also quantized, the spectrum is  discrete and shifted by a constant value proportional to $k_+$. Globally, the M2-branes considered are described in terms of symplectic torus bundles over a torus with a monodromy contained in $SL(2,Z)$ and with a topologically non trivial $U(1)$ connection \cite {mpgm3}.
In \cite{mpgm10} it was proved that these structures generate a twisted torus bundle,  where the base manifold is given by the worldvolume Riemann surface $\Sigma$, the fiber is a twisted torus $\mathbb{T}^3$ and the structure group are the 2-torus symplectomorphisms, $Symp(T^2)$. The consistence of the global description and all of the details are discussed in \cite{mpgm10}. 
%%%%%%%%%%%%%%%%%%%%%%%%%%%%%%%%%%%%%%%%%%%%%%%%%%%%%%%%%%%%%%%%%%%%%%%%%%%%%%%%%%%%%%%%%%%%%%%%%%%%%%%%%%%%%%%%%%%%%%%

\section{CM2-brane on Twisted Torus Bundle}
\label{Sec3}
We now obtain a Hamiltonian formulation of the M2-brane on a quantized constant three-form on the same target space $M_9\times T^2$, in which the transverse components of the three-form become explicit. To distinguish it from the previous case, we will refer to it as CM2. Let us note that we can also decompose $C_a$ (\ref{CaLCG}) to avoid the nonphysical degrees of freedom  as $C_a=C_a^{(1)} + C_a^{(2)}$  with
\begin{eqnarray}
C_a^{(1)} &=&-\epsilon^{uv}\partial_u X^- \partial_v X^b C_{-ab} \quad , \quad C_a^{(2)}= \frac{1}{2}\epsilon^{uv}\partial_u X^a \partial_v X^b C_{abc}
\end{eqnarray}
Therefore, instead of the previous canonical transformation (\ref{TransCanonica}), it is enough to assume
\begin{eqnarray}\label{TransCanonicaaa}
\widehat{P}_a &=& P_a - C_a^{(1)} \quad , \quad \widehat{P}_- = P_- - C_-
\end{eqnarray}
to obtain the Hamiltonian in terms of its physical degrees of freedom. In fact, it can be verified that it also constitutes a consistent canonical transformation of the phase space variables, preserving the kinetic term and the Poisson brackets of the theory. The associated $11D$ LCG Hamiltonian becomes
\begin{eqnarray}\label{CM2}
\mathcal{H}_{CAN}=\left[\frac{1}{2}\frac{( \widehat{P}_a - C_a^{(2)})^2}{\sqrt{W}} + \frac{1}{4}\sqrt{W}\left\lbrace X^a,X^b \right\rbrace^2 - \sqrt{W} \bar{\theta}\Gamma^-\Gamma_a \left\lbrace X^a,\theta \right\rbrace  - C_+\right]
\end{eqnarray}
where in distinction to the case previously analyzed (\ref{hamiltoniancpm}) the component $C_a^{(2)}$ is present carrying the information of the $C_{abc}$ components of the three-form, with $a=1,\dots,9$ denoting the transverse components. The trivial dynamics of $\widehat{P}_-$ allow us to set $\widehat{P}_-=\widehat{P}_-^0\sqrt{W}$. If now we perform a toroidal compactification and impose the flux condition (\ref{fluxpullback}) that acts as an extra constraint, then the LCG CM2-brane Hamiltonian with $C_{\pm}$ fluxes becomes 
\begin{eqnarray}\label{HCMIM2}
H_{CM2} &=& \int d^2 \sigma \sqrt{W} \left\lbrace
\frac{1}{2}\left(\frac{\widehat{P}_m-C_m^{(2)}}{\sqrt{W}}\right)^2 + \frac{1}{2}\left(\frac{\widehat{P}_r-C_r^{(2)}}{\sqrt{W}}\right)^2 + \frac{1}{4} \left\lbrace X^m,X^n \right\rbrace^2 +  \frac{1}{4}(\widehat{F}^{rs})^2\right.  \nonumber \\
&+& \left.   \frac{1}{2}\left( \mathcal{D}_r X^m\right)^2 +\frac{1}{4} (\mathcal{F}^{rs})^2 -\bar{\theta}\Gamma^-\Gamma_r\mathcal{D}_r\theta - \bar{\theta}\Gamma^-\Gamma_m\left\lbrace X^m,\theta\right\rbrace-C_{+}    \right\rbrace 
\end{eqnarray}
where
\begin{eqnarray}
\small
C_m^{(2)} &=& \frac{\epsilon^{uv} }{2}\left[ \partial_u X^{\bar{n}} \partial_v X^n C_{m\bar{n}n} + 2\partial_u X^n \partial_v X^r C_{mnr}+\partial_u X^r \partial_v X^s C_{mrs}\right] \\
C_r^{(2)} &=& \frac{\epsilon^{uv}}{2}\left[\partial_u X^m \partial_v X^n C_{rmn} + 2\partial_u X^m \partial_v X^s C_{rms}\right]
\end{eqnarray}
subject to the same APD constraints (\ref{APDconstraints1}) and (\ref{APDconstraints2}). The only nontrivial contribution on the $C_+$ term is due to $C_{+rs}$ components. On the Hamiltonian (\ref{HCMIM2}) appears explicitly the contribution of the transverse components of the supergravity three-form $C_{abc}$ with $a=(m,r)$,  
in distinction with the Hamiltonian of an M2-brane with $C_\pm$ fluxes (\ref{hamiltoniancpm}). 
However, a redefinition of the phase space variables based on the following canonical transformation reveals that both Hamiltonians, i.e. the CM2-brane and the M2-brane with $C_{\pm}$ fluxes, are equivalent.
 \begin{eqnarray}\label{ct3}\widetilde{P}_m = \widehat{P}_m - C_m^{(2)},\quad \widetilde{P}_r = \widehat{P}_r - C_r^{(2)}.
\end{eqnarray}
 Consequently, the two theories previously obtained, and this last new one must share the same qualitative spectral properties of discreteness of the supersymmetric spectrum. This result is not obvious from a direct examination of (\ref{HCMIM2})  at the regularized level. This is due to the coupling of the scalar field's pullback and the canonical momenta. Indeed, the sufficiency criteria for the discreteness of the supersymmetric spectrum \cite{Boulton} can not be directly applicable. Because the sector of the M2-brane with central charges and the one with a quantized constant three form represented by the CM2-brane can also be connected by canonical transformations, this can be interpreted as additional evidence of a duality between both previously unconnected sectors, realized at the Hamiltonian and mass operator levels.  Globally, the CM2-brane can also be described in terms of a Twisted Torus Bundle with monodromy contained in $SL(2,Z)$.
%%%%%%%%%%%%%%%%%%%%%%%%%%%%%%%%%%%%
\section{D-branes from nontrivial M2-branes}
In this section, we obtain the D-brane description from the CM2-brane Hamiltonian formulation. To this end, we use the scalar/vector duality and we will refer to these sectors as duals in the aforementioned sense. In this paper we choose the simplest possible background where the nontriviality of the M2-brane becomes manifest. Their formulation is done in the LCG since the toroidally nontrivial M2-branes have been formulated in this context in order to establish their spectral properties. The LCG fixing commutes with the dualization procedure. Previous formulations of the  LCG D2-brane Hamiltonian in ten dimensions in the absence of RR and NSNS background fields were done in \cite{Manvelyan1,Manvelyan2,Lee}. For the case of toroidally compactified target spaces, the Hamiltonian was obtained in \cite{Ovalle1,Ovalle2}. The virtue of working with the CM2-brane formulation is that the presence of the $C_{abc}$ transverse components, under dualization, generates  the B-field coupling inside the Born Infeld action, explicitly.  From the CM2-brane dual formulation, the duals of the other nontrivial M2-branes cases previously discussed, can be obtained. The 11D flux condition is dualized to produce 10D RR and NSNS flux conditions. When the D2-brane theories contain fluxes associated with the presence of RR and/or NSNS charges, they have been conjectured to admit a description in terms of D-brane bound states.
%%%%%%%%%%%%%%%%%%%%%%%%%%%%%%%%%%%%%%%%%%%%%%%%%%%%%%%%%%%%%%%%%%%%%%%%%%%%%%%%%%%%%%%%%%% 
\subsection{D-brane description of CM2-brane.} 
The action corresponding to the CM2-brane with $C_{\pm}$ fluxes subject to APD (\ref{APDconstraints1}) and (\ref{APDconstraints2}), can be obtained from (\ref{HCMIM2}) by a Legendre transformation. Following \cite{Townsend4}, let us consider a compactification on an extra circle, $X^m=(X^\alpha,X^9)$ with $\alpha=3,\dots,8$ and isolate the contribution of $X^9$ on the action. If we promote $L=dX^9$ to an independent one-form on the worldvolume by adding a topological term $AdL$, such that $dL=0$ we have
\begin{eqnarray}\label{dualaction}
S &=& S_0 + \int d^3\xi\left[ \frac{1}{2}\sqrt{W}\left(L_0+\frac{\epsilon^{uv}}{\sqrt{W}}\partial_u\Lambda L_v\right)^2 - \frac{1}{2}\sqrt{W}\left(\frac{\epsilon^{uv}}{\sqrt{W}}\partial_u X^\alpha L_v\right)^2\right. \nonumber \\
&-& \left. \frac{1}{2}\sqrt{W}\left(L_0+\frac{\epsilon^{uv}}{\sqrt{W}}\partial_u\Lambda L_v\right)\widehat{F}^{rs} B_{rs} - \frac{1}{2}\sqrt{W}\left(\frac{\epsilon^{uv}}{\sqrt{W}}\partial_u X^r L_v\right)^2 \right.\nonumber \\
&-& \left. \sqrt{W}\left(L_0+\frac{\epsilon^{uv}}{\sqrt{W}}\partial_u\Lambda L_v\right)(\mathcal{F}^{rs} B_{rs})-\frac{1}{2}\sqrt{W}\left(\frac{\epsilon^{uv}}{\sqrt{W}}\partial_u X^s L_v B_{rs}\right)^2\right. \nonumber \\´
&-& \left. \sqrt{W}\left(\frac{\epsilon^{uv}}{\sqrt{W}}L_u X^r B_{+r}\right)  \right]-  \frac{1}{2}\int d^3\xi \left\lbrace \epsilon^{uv}L_0F_{uv} + 2\epsilon^{uv}L_uF_{v0} \right\rbrace
\end{eqnarray}
where $L_0=\Dot{X}^9$, $L_u=\partial_uX^9$, $F=dA$, $\Lambda$ is the Lagrange multiplier and $B_{rs}=C_{9sr}$ are the components of the NSNS background field on the compact sector.  The term denoted as $S_0$ corresponds to
\begin{eqnarray}
S_0&=& \int d^3\sigma \left\lbrace \frac{1}{2}\frac{\widehat{P}_\alpha \widehat{P}^\alpha}{\sqrt{W}} + \frac{1}{2}\frac{\widehat{P}_r \widehat{P}^r}{\sqrt{W}} - \frac{1}{4} \left\lbrace X^\alpha,X^\beta \right\rbrace^2 -  \frac{1}{4}(\widehat{F}^{rs})^2\right.  \nonumber \\
&-& \left.   \frac{1}{2}\left( \mathcal{D}_r X^\alpha\right)^2 -\frac{1}{4} (\mathcal{F}^{rs})^2+C_{+} \right\rbrace
\end{eqnarray}
For simplicity we have considered that $C_{\pm rs}^{(10)}$, $B_{\pm r}$ and $B_{rs}$ are the only nontrivial but constant components of the background fields in 10D. However, is easy to verify that $B_{-r}$ and $C_{-rs}$, do not appear explicitly in the Hamiltonian.\newline
 From the equations of motion we get
\begin{eqnarray}
L_0&=&\left[\frac{\epsilon^{uv}F_{uv}}{2\sqrt{W}} - \left\lbrace \Lambda,X^9\right\rbrace + \frac{1}{2}\widehat{F}^{rs} B_{rs} + \frac{1}{2}\mathcal{F}^{rs}B_{rs} \right]  \\
 L_\omega &=& \sqrt{W}\frac{\Tilde{\gamma}_{\omega v}}{\Tilde{\gamma}} \left[ \frac{1}{2}\frac{\epsilon^{\rho\sigma}F_{\rho\sigma}}{\sqrt{W}}\epsilon^{v\bar{v}}\partial_{\bar{v}}\Lambda - \epsilon^{v\bar{v}}F_{0\bar{v}} - \epsilon^{v\bar{v}}\partial_{\bar{v}} X^r B_{+r}\right]
\end{eqnarray}
 By inserting these expressions in the previous action and performing a Lagrange transformation the corresponding Hamiltonian in $M_8\times T^2$, it becomes
\begin{eqnarray}\label{HCMIM22}
\mathcal{H}_{dual} &=& \frac{1}{2} \frac{\widehat{P}_\alpha  \widehat{P}^\alpha}{\sqrt{W}} + \frac{1}{2}\frac{(\widehat{P}_r - B_r)^2}{\sqrt{W}} + \frac{1}{2}\frac{\Pi^u\Pi^v \widetilde{\gamma}_{uv}}{\sqrt{W}} + \frac{G}{2\sqrt{W}} + \sqrt{W}\frac{1}{4}(\widehat{F}^{rs})^2  \nonumber \\
&+& \frac{1}{4}\sqrt{W}(\mathcal{F}^{rs})^2+ \frac{1}{2}\sqrt{W}(\mathcal{D}_r X^\alpha)^2 - C_{+}^{(10)} - B_+
\end{eqnarray}
where\footnote{By $\gamma$ we denote 
$\gamma=det(\gamma_{uv})= det(\partial_uX^\alpha\partial_vX_\alpha)$ and $\widetilde{\gamma}_{uv}=\gamma_{uv}+\partial_uX^r\partial_v X_r$} 
$G=\gamma + \mathcal{F}^{DBI}$ $, \quad B_r=\Pi^u\partial_u X^s B_{rs}, \quad B_+=\Pi^u\partial_u X^s B_{+s}, \quad \Pi^u=\epsilon^{uv}L_v$,
  and $\mathcal{F}^{DBI}= det(F_{uv} + B_{uv})$ being
\begin{eqnarray}
\small
\label{Fcursiva}
\mathcal{F}^{DBI} &=& F + b_2(\sqrt{W})^2\left[\frac{1}{2}b_2\left(\widehat{F}^{rs}\epsilon_{rs} + \mathcal{F}^{rs}\epsilon_{rs}\right)^2 + (*F)\left( \widehat{F}^{rs} + \mathcal{F}^{rs}\right)\epsilon_{rs}\right]
\end{eqnarray}
with $B_{uv}=\frac{1}{2}\partial_u X^r \partial_v X^s B_{rs}$, $B_{rs}=b_2 \epsilon_{rs}$ and where  the RR ten dimensional $C_+^{(10)}$ has the same form of its 11D dual counterpart $C_+$ (\ref{CaLCG}).
The LGG Hamiltonian (\ref{HCMIM22}) is subject to Gauss law and local and  global APD constraints,
\begin{eqnarray}\label{D2constraints}
\partial_u \Pi^u &\approx& 0. \label{Gaussconstraint}\\
\left\{ (\sqrt{W})^{-1}\widehat{P_\alpha} , X^\alpha\right\} + \mathcal{D}_r\left( (\sqrt{W})^{-1}\widehat{P}_r\right) + \epsilon^{uv}\partial_u\left[ \frac{\Pi^\omega F_{v\omega}}{\sqrt{W}} \right] &\approx& 0 \label{APD1}\\
 \oint_{C_S}\left[\frac{\widehat{P}_\alpha \partial_u X^\alpha}{\sqrt{W}} + \frac{\widehat{P}_r \partial_u X^r}{\sqrt{W}} + \frac{\Pi^\omega F_{u\omega}}{\sqrt{W}}\right]d^u\sigma &\approx& 0. \label{APD2}
\end{eqnarray}
The M2-brane flux conditions are used to derive the D-brane flux conditions. As \cite{Townsend4} originally observed in the covariant formulation, if one performs in the LCG, the dualization to the WZ term in eleven dimensions, the background fields become
$C_{\pm} = C_{\pm}^{(10)} + B_{\pm}, \label{C+11}$ and $C_a = C_a^{(10)} - B_a.$
with  $B_-=\Pi^u\partial_u X^s B_{-s}$. Gauge invariance of the two-form allows us to fix $B_{-s}=0$. Hence, it can be seen that the $C_{\pm}$ quantization condition in $D=11$, implies in the dual action, 
 \begin{eqnarray}\label{quantization}
\int_{\widetilde{\Sigma}} C^{(10)}_{\pm}=k_{\pm}^{(10)},\quad  \int_{\widetilde{\Sigma}} B_+ = b_+,\quad \int_{\widetilde{\Sigma}} B_2 = b_2 \int_{\widetilde{\Sigma}} \widehat{F} =b_2 n
\end{eqnarray}
 where the components $C_{\pm rs}^{(10)}=c_{\pm}^{(10)}\epsilon_{rs}$ and $B_{rs}=b_2\epsilon_{rs}$  with  $k_{\pm}^{(10)}=n c_{\pm}^{(10)}$  such that they satisfy $k_{\pm}=k_{\pm}^{(10)}+b_+$.
The components of the 10D RR three-form are constant and they generate 2-form fluxes whose pull-back on the worldvolume of the D2-brane, in analogy with the M2-brane analysis, are associated with a topologically nontrivial U(1) curvature. There is an extra contribution to the B-field that appears in the DBI term, associated with the pull-back of the Kalb-Ramond  field $B_2=\frac{1}{2}B_{rs}dX^r\wedge dX^s$. This contribution -since the coefficient is also constant -  also generates a two form flux condition on the worldvolume. At first sight, it could seem that there is not a flux quantization condition acting on it. However, due to the fact that the coefficient is constant, the worldvolume flux condition is automatically guaranteed by the central charge condition induced by the $C_{\pm}$ quantization of the M2-brane. 
This U(1) worldvolume monopole condition found in \cite{Restuccia} associated with a first nontrivial Chern class given by $k_{\pm}+b_2$ contributes to the Hamiltonian with a nontrivial curvature $\widehat{F}$ associated with a nontrivial $U(1)$ connection $\widehat{A}$ under symplectomorphisms transformations. As a consequence, the dual Hamiltonian of the CM2-brane on $M_9^{LCG}\times T^2$ can be understood as a D2-brane with 2-form fluxes given by (\ref{quantization}). See that in the same way that the flux constraint  for the nontrivial M2-brane  generates a symplectic gauge field with an associated symplectic field strength ($\mathcal{F}$), it also happens in the D2-brane associated to the CM2-brane with $C_{\pm}$ fluxes. 

As previously stated, because the D2-brane has RR three-forms and NSNS two-forms, both of which generate 2-form fluxes, an open question is whether it can be described in terms of Dp bound states. This 'dual' D-brane theory inherits from the nontrivial M2-brane, the same type of $U(1)$ topologically nontrivial gauge field. In fact, this condition was discovered in \cite{Restuccia} an it represents a generalization of the Dirac monopole condition to Riemann surfaces of  genus $g\ge 1$ \cite{Restuccia2}. This quantization condition acts as an extra constraint on the  Hamiltonian, generating new dynamical fields, as we will see in more detail in section 5.   In \cite{mpgm11}  the spectral properties of the $SU(N)$ regularized M2-brane with central charge and the $0+1$ dimensionally reduced D2-D0 bound state were compared.  Though at first sight both models seem quite similar and carry a RR charge, they are associated with different monopole conditions and their spectra are completely different. In \cite{mpgm11} it was rigorously shown that the regularized M2-brane theory with central charges has a purely discrete supersymmetric spectrum, while the dimensionally reduced D2-D0 has a continuous spectrum bounded by below by the $U(N)$ monopole energy contribution to infinity, as originally conjectured by \cite{Witten4}. On the D-brane side, the CM2 dual (\ref{HCMIM22}) has a DBI curvature $F$ that is not quantized in contrast to D2-D0 bound states. This analysis, however, does not rule out the possibility of expressing it in terms of more complicated bound state constructions, such as in  \cite{mpgm20}. 
%%%%%%%%%%%%%%%%%%%%%%%%%%%%%%%%%%%%%
\section{D-brane description: new features}

In this section, we are going to emphasize the physical implications of the nontrivial D2-branes. On one hand, we have seen that the quantization condition on the M2-branes implies the existence of nontrivial quantization conditions on the constant RR and NSNS background fields of the D2-brane. These particular quantization conditions generate worldvolume 2-form flux fields. According to \cite{Restuccia2} this automatically implies that the associated D2-brane -which we will refer to as nontrivial D2-branes in the following - must have a monopole charge given by the units of fluxes turned on the worldvolume. Secondly, in the cases when $B_{rs}$ is quantized, the $\mathcal{F}^{U(1)}$ appearing in the Dirac Born-Infeld action is not topologically trivial. 
We will see that the nontrivial D2-brane posses different degrees of freedom, hence dynamical fields, than those associated to a usual toroidally compactified D2-brane. On top of  the embedding scalar fields $X^m$ and the standard $U(1)$ DBI gauge field $A_u$, it appears a new singlevalued symplectic gauge field $\mathcal{A}_r$ and new nontrivial symmetries on the worldvolume.

 %%%%%%%%%%%%%%%%%%%%%%%%%%%%%%%%%%%%%%%%%%%%%%
 \subsection{New gauge Symmetries}
The dual description of a D2-brane contains two characteristic  symmetries related to the worldvolume $\widetilde{\Sigma}$: DBI U(1) symmetry related to the Gauss constraint and the symplectomorphisms. It is easy to see that  $A_u$ on $\widetilde{\Sigma}$  transforms under the Gauss constraint as a U(1) gauge field
\begin{eqnarray}
\delta A_u = \partial_u \Omega, \label{U(1)DBI}
\end{eqnarray}
 where $\Omega=-\Lambda$ is the parameter of the transformation. The dynamical variables $(X^m,\mathcal{A}^r,A_u)$ also transform under the APD constraint. In fact, concerning to the symplectomorphisms connected to the identity, any functional $O$ of the canonical variables transform locally under ${Symp}_0(T^2)$ as \cite{Restuccia,Restuccia3} 
\begin{eqnarray}
\delta O &=& \left\lbrace O, <\xi \phi>\right\rbrace_{PB}
\end{eqnarray}
where $\phi$ is the local constraint (\ref{APD1}) and for example, for the first term in (\ref{APD1}) we have that $< >$ denotes
\begin{eqnarray}
 <d\xi\wedge \frac{P_m}{\sqrt{w}}dX^m> \equiv \int_{\widetilde{\Sigma}} d^2 \sigma \sqrt{w} \left(d\xi\wedge \frac{P_m}{\sqrt{w}}dX^m\right)
\end{eqnarray}
with $\xi=\xi(\sigma,\tau)$ 
 the continuous parameter that contains both, the local and global parameters, associated to the $Symp_0(\Sigma)$ transformations \cite{Restuccia4}. Therefore, under symplectomorphisms connected to the identity we have that
\begin{eqnarray}
\delta_{\xi} X^m &=&  \left\lbrace \xi, X^m \right\rbrace \\
\delta_{\xi} A_u &=&  \xi^v F_{vu} = \left\lbrace \xi, A_u \right\rbrace + \xi^v\partial_u A_v, 
\end{eqnarray}
where $m$ runs over the compact and non-compact indices, $m=(r,\alpha)$ with $r=1,2$ and $\alpha=3,\dots,8$.

If we now examine the maps of the compact sector under symplectomorphism transformations, taking into account the worldvolume flux condition and the Hodge decomposition, we can split them in the following manner, which is similar to the study performed in
\cite{mpgm10}
\begin{eqnarray}\label{eqn 4.6}
\delta_{\xi} X^r =\left\lbrace \xi, X^r\right\rbrace \,= \delta X_h^r + \delta \mathcal{A}^r .
\end{eqnarray}
Let us remark that the single-valued function of the embedding map  $\mathcal{A}^r$ defines an associated one form $d\mathcal{A}$ -with $r=1,2$  the index running over the directions of the $T^2$- which transforms as a symplectic connection in contrast to the DBI  U(1) connection $A_u$ with $u=1,2$ defined over $\widetilde{\Sigma}$.

\subsubsection{Symplectic Gauge Symmetry}
 Similarly to the nontrivial M2-branes considered, we define a class of maps whose associated one-forms $dX_h$ are expressed in terms of the harmonic basis such that under infinitesimal transformations of the type $\delta X_h^r=\{\nu, X_h^r\}$ with $\nu$ an infinitesimal parameter, a curvature $\widehat{F}=\epsilon_{rs}dX_h^r\wedge dX_h^s$ with $r,s,=1,2$ is preserved.
When the symplectomorphisms transformations (\ref{eqn 4.6}) are realized as follows,
\begin{equation}
\delta [X_h^r] = 0 \quad \text{and} \quad \delta \mathcal{A}^r = \mathcal{D}_r\xi \, ,
 \end{equation}
the multivalued $X_h^r$ becomes inert under the transformation and all of the transformation is realized by the $\mathcal{A}^r$ scalar field. The transformation law for $d\mathcal{A}^r$, corresponds to a symplectic one-form connection over  $\widetilde{\Sigma}$.   The associated one-form transforms as a symplectic gauge field. Its associated symplectic curvature is  $\mathcal{F}=\mathcal{D}A+\{A,A\}$ is equivalent to (\ref{simplectica}), which is topologically trivial

It defines a symplectic covariant derivative $\mathcal{D}\bullet=D\bullet+\{\mathcal{A},\bullet\}$
whose transformation, preserves the transformation law of its argument under symplectomorphisms.

In the context of toroidally nontrivial M2-branes, a similar symplectic gauge field, its associated covariant derivative, and its curvature had previously been identified. There and here, its origin is found in the action of the specific  constant 2-form fluxes acting over the worldvolume, which act as an extra constraint on the Hamiltonian imposing a restriction on the embedding maps. 
%%%%%%%%%%%%%%%%%%%%%%%%%%%%%%%%%%%%%%%%%%%%%%%%%%%%%%%%%%%%%%%%%%%%%%%%%%%%%%%%%%%%%
\subsubsection{U(1) Gauge Symmetries}
 Any D-brane wordlvolume action contains a  characteristic DBI U(1)  connection $A = A_u d\sigma^u$, where the transformation of $A_u$ under the Gauss constraint is given by (\ref{U(1)DBI}). 2-form fluxes, on the other hand, naturally induce a topologically nontrivial $U(1)$ over their worldvolume, though the precise effect on the Hamiltonian depends on the type of flux and its origin. Let us notice that the flux conditions on $C_{\pm}$ and $B_2$ given by (\ref{quantization}) imply the existence of a non trivial U(1) one-form $\widehat{A}=\frac{1}{2}\epsilon_{rs}X_h^rdX_h^s$, with an associated curvature $\widehat{F}=d\widehat{A}$ defined as in \cite{mpgm10}. 
In the case of constant 2-form fluxes, the associated $U(1)$ dynamical gauge symmetry is related to the one-form defined in terms of scalar fields via a nontrivial transformation via symplectomorphisms. We recall that this symmetry is generated by the APD constraint (\ref{eqn 4.6}) over the D2-brane worldvolume $\widetilde{\Sigma}$ by the following transformation
  consistent with (\ref{eqn 4.6})
\begin{equation}\label{transformacion1}
\delta [X_h^r] = \left\lbrace \xi, [X_h^r] 
\right\rbrace \quad \text{and} \quad \delta \mathcal{A}^r = \left\lbrace \xi, \mathcal{A}^r  \right\rbrace \,.   
 \end{equation}
 In fact, we also may define  $A_{\pm}=k_{\pm}\widehat{A}$ and $A_B=b_2\widehat{A}$, the non trivial U(1) connections reminiscent of the quantization conditions over the RR constant three-form and the NSNS constant 2-form, with  $F^{(wv)}=k_{\pm}\widehat{F}$ and $B_2=b_2 \widehat{F}$, respectively. They contribute to the total amount of 2-form flux defined on  the D2-worldvolume. In fact, it can be checked that $\widehat{A}$ does transform as a U(1) connection under symplectomorphisms
\begin{eqnarray}
\delta \widehat{A} =d \eta \,, \quad\textrm{with a parameter\quad } \eta = -\frac{\epsilon^{uv}}{\sqrt{w}}\partial_v\xi \left(\widehat{A}_u\right)
-\xi\left(\star\widehat{F}\right), 
\end{eqnarray}
where
\begin{eqnarray} 
\frac{1}{2\pi}\int_\Sigma \widehat{F} = n.
\end{eqnarray}
%and
%\begin{eqnarray}
%\star \widehat{F} = \frac{\epsilon^{uv} F_{uv}}{2\sqrt{W}} = n,
%\end{eqnarray}
As a result, the nontrivial one-form connection on the worldvolume is  $A_{\pm}+A_{B}=(b_2+k_{\pm})\widehat{A}=k\widehat{A}$. Due to the constant nature of the NSNS background field $B_2$, its pullback to the worldvolume corresponds to the field strength of a one-form -which also transforms as a non trivial U(1) connection under symplectomorphisms-. 
Closely following \cite{mpgm10} it is possible to define a new dynamical   topologically trivial $U(1)$ one-form $\mathcal{A}_G$ defined in terms of scalar embedding maps as follows,
\begin{eqnarray}
\mathcal{A}_G = \frac{1}{2}\epsilon_{rs}(\mathcal{A}^rdX_h^s-\mathcal{A}^sdX_h^r+\mathcal{A}^rd\mathcal{A}^s).
\end{eqnarray}
 It transforms as a $U(1)$ connection under the symplectomorphisms specified by (\ref{transformacion1})
\begin{eqnarray}
\delta \mathcal{A}_G &=& d \widetilde \eta  \,, \quad \widetilde \eta\equiv \left( -\frac{\epsilon^{uv}}{\sqrt\omega} \partial_{v}\xi (\frac{1}{2}\epsilon_{rs}\mathcal{A}^r\partial_{u}X_h^s) -\xi \left(* \widehat F \right)\right) \label{ACBcursiva6}
\end{eqnarray}
with an associated curvature labelled $ \mathcal{F}_{G}=d\mathcal{A}$ topologically trivial.
A particular property of this curvature $\mathcal{F}_G=d\mathcal{A}_G$ is that it coincides with the symplectic curvature (\ref{Fcursiva}). In fact, it was shown in the context of the M2-brane in \cite{mpgm10}. Furthermore, both structures are consistent with the irreducible wrapping condition. Here, the same analysis remains valid. 
 %We may also consider a case where $k_{\pm}+k_B=k$, in which case the sum of both conditions generates a global curvature proportional that the central charge condition $kn$.\footnote{The worldvolume flux condition can be also called central charge condition by defining $\widetilde{n}=kn$ over the worldvolume induced by the pullback of the two form RR and NSNS background fluxes. We have followed in this paper the original definition. }
Hence, it is possible to define a more general $U(1)$ connection $\widetilde{\mathbb{A}}$ in terms of $\widehat{A}$ and $\mathcal{A}_G$ \begin{eqnarray}\label{Agordita}
\widetilde{\mathbb{A}} = \widehat{A} + \beta \mathcal{A}_G
\end{eqnarray}
with $\beta$ a real scalar.

 The connection is 
defined on the same nontrivial principal bundle characterized by the first Chern class $n$. It has an associated curvature $\mathbb{F} = d\mathbb{A}$ that satisfies,
\begin{eqnarray}
\frac{1}{2\pi}\int_\Sigma \mathbb{F}= n\ne0
\end{eqnarray}
with
 \begin{eqnarray}
\mathbb{F}=\widehat{F}+\beta\mathcal{F} 
\end{eqnarray}
 As a result, the D-brane description of the non trivial M2-brane duals, discussed here, has two U(1) gauge symmetries, whose curvatures are respectively $F$,  $\mathbb{F}$ in distinction with the usual case.
Indeed, the $\mathcal{F}^{DBI}$ can be re-expressed in terms of this new dynamical topologically nontrivial $U(1)$ as,
\begin{eqnarray}
\small
\label{Fcursiva2}
\mathcal{F}^{DBI} &=& F + \frac{1}{2}b_2(\sqrt{W})^2\left[b_2\left(\mathbb{F}^{rs}\epsilon_{rs}\right)^2 + 2(*F)\mathbb{F}^{rs}\epsilon_{rs}\right]
\end{eqnarray}
This naturally reinforces the idea that $H_{dual}$ could be better described by a bound state of D-branes.
%%%%%%%%%%%%%%%%%%%%%%%%%%%%%%%%%%%%%%%%%%%%
\subsection{Twisted torus Bundle description} The characteristics of the aforementioned nontrivial D2-branes introduce new aspects into the bundle description that we characterize.  The topologically non trivial part of a standard toroidal D2-brane is characterized by two independent fibers: a principal torus bundle defined over its worldvolume $\widetilde{\Sigma}$ and a  $U(1)$ fiber associated with the DBI contribution.  In the presence of a two-form fluxes that are not associated with the DBI gauge field, a nontrivial $U(1)$ fiber is added. As we will see, in the case we are considering, there is an extra relationship between the fibers which changes the overall construction.

In order to introduce the global description of the non trivial M2-brane duals, let us recall that the  D2-branes are formulated on $M_8\times T^2$ with a foliation on the worldvolume, such that $\widetilde{\Sigma}$ is a Riemann surface of genus one related to the spatial directions of the worldvolume.  It contains a topologically trivial $U(1)$ principle bundle defined on it associated to the DBI contribution. Flux condition induced by the constant quantized background condition implies, through the  Weyl theorem, the existence of a U(1) principal bundle over $\widetilde{\Sigma}$ 
\begin{eqnarray}
U(1)\rightarrow E' \rightarrow \widetilde{\Sigma}.
\end{eqnarray} 
Now, because the D2-branes are toroidally compactified but the flux acts as an extra constraint,in close analogy with \cite{mpgm3}, it admits a symplectic torus bundle description with monodromy in $SL(2,Z)$ defined as
\begin{eqnarray}
T^2 \rightarrow E \rightarrow \widetilde{\Sigma} \, ,\hspace{0.5cm} G=Symp(T^2)
\end{eqnarray}
with $T^2$ is the compact part of the fiber, $\widetilde{\Sigma}$ is the base, $E$ is the total space and $G$ is the structure group of the fiber bundle. 
 It is possible to define a Maurer-Cartan structure between the flux condition and the torus of the fiber, such that they defined a twisted three-torus. In fact, we may define three global one-forms
 \begin{eqnarray}
e^1&=&d\widetilde{X}^1 , \\
e^2&=&d\widetilde{X}^2 , \\
e^3&=&dy + k\widetilde{X}^1d\widetilde{X}^2
\end{eqnarray}
where $(\widetilde{X}^1,\widetilde{X}^2)\in T^2$ and $y$ is a coordinate on the $S^1$ associated with the nontrivial $U(1)$ fiber, such that the global one-forms $e_1$, $e_2$ y $e_3$ satisfies the structure equation
\begin{eqnarray}
de^3=f^3_{12}e^1\wedge e^2
\end{eqnarray}
with $f^3_{12}=k$ known as Maurer-Cartan. 
 
In consequence, as in  \cite{mpgm10} the two fibers form  terms of a Twisted Torus Bundle
\begin{eqnarray}
(T^2)^{U(1)} \rightarrow E \rightarrow \Sigma \, ,\hspace{0.5cm} G=Symp(T^2)
\end{eqnarray}
where $(T^2)^{U(1)}$ denotes a 2-torus with a U(1) monopole connection over it, equivalent to a twisted 3-torus.

Nontrivial D2-branes with worldvolume and background fluxes on $M_8\times T^2$ can thus be described geometrically as twisted torus bundles with monodromy in $SL(2,Z)$ and an extra $U(1)$ trivial principal bundle associated with the DBI gauge symmetry. They are duals of nontrivial M2-branes on $M_9\times T^2$  formulated on a twisted torus bundle with monodromy in $SL(2,Z)$.

%%%%%%%%%%%%%%%%%%%%%%%%%%%%%%%%%%%%%%%%%%%%%%
\section{ D2-branes on a RR and NSNS background with fluxes.} In this section, we will show how, when certain 2-form fluxes are included, the LCG Hamiltonian of (\ref{HCMIM22}) can be also obtained directly from the D2-brane. It is important to note, however, that this formulation differs from a usual D2-brane with RR and NSNS background fields. To demonstrate it, we illustrate its direct construction, emphasizing its key points. We will investigate the LCG D2-brane in the presence of constant RR and NSNS background subject to quantization conditions, because it will be compared to the LCG formulation of the M2-brane on a general quantized constant supergravity three-form $C _3$, which is relevant for M2-brane spectral characterization.
By means of the dimensional reduction and the scalar/vector duality we have obtained the D-brane related to the nontrivial CM2-brane. 
%The Hamiltonian is given by (\ref{}) and it is related to a D2-brane with 2-form flux conditions on the worldvolume. In order to reproduce this Hamiltonian from the covariant DBI action, we have to solve an issue related to the presence of nonphysical degrees of freedom in the LCG Hamiltonian, before the imposition of the nontrivial flux condition.

The covariant formulation of the Dp-brane action in the presence of RR and NSNS background was originally found in \cite{Townsend5}. The LCG D2-brane Hamiltonian on a flat background was obtained in \cite{Manvelyan1, Lee}. However, in these works, the coupling with the RR and NSNS background fields was not considered. As previously mentioned in relation to the M2-brane \cite{mpgm6}, the difficulty of its formulation relies on the proper handling of the non-physical degrees of freedom. It requires the obtention of proper canonical transformations for the D2-brane theory that allow us to eliminate the $X^-$ dependence without introducing nonlocalities.

Consider the Lagrangian density of a D2-brane on a 10D Minkowski spacetime, coupled to the RR 3-form and NSNS 2-form background fields,
\begin{eqnarray}
\mathcal{L}_{D2}=-\sqrt{-\Bar{G}} -\frac{1}{3!}\epsilon^{ijk}\partial_iX^\mu \partial_jX^\nu\partial_kX^\rho C^{(10)}_{\mu\nu\rho}, 
\end{eqnarray}
with $X^{\mu}(\sigma^1,\sigma^2,\tau)$ the D2-brane embedding maps from the worldvolume $\tilde{\Sigma}$ in the $10D$ target-space $M_{10}$ labelled by $\mu,\nu,\rho=0,\dots,9$ and $i,j,k=0,1,2$ denoting the worldvolume indices.  Let us begin by defining the  generalized metric,
\begin{eqnarray}
\Bar{G} &=& det(\gamma_{ij}+\mathcal{F}_{ij}^{DBI}),
\end{eqnarray} 
  in terms of the worldvolume induced metric $\gamma_{ij}=\partial_i X^\mu \partial_j X_\mu$  and  on the curvature 
\begin{eqnarray}
\mathcal{F}^{DBI}_{ij} &=& F_{ij}+ B_{ij}.
\end{eqnarray}
It is defined in terms of the  $U(1)$ Born-Infeld field strength $F_{ij}=\partial_iA_j-\partial_jA_i$ and a B-field which is the pullback of the NS-NS two-form background to the worldvolume $B_{ij}= \partial_iX^\mu \partial_j X^\nu B_{\mu\nu}$. 
It is straightforward to see that the LCG Lagrangian density is given by
\begin{eqnarray} \label{LCG L RR NSNS}
\mathcal{L}&=& -\sqrt{\triangle G} + C^{(10)}_{+-} + C^{(10)}_+ + C^{(10)}_M\partial_0X^M + C^{(10)}_-\partial_0X^-,
\end{eqnarray}
where $\Bar{G}=-\triangle{G}$ with $\triangle=-G_{00}+G_{0u}G^{uv}G_{v0}$, $G=detG_{uv}$ and $u,v=1,2$ being the spatial worldvolume indices. Following the notation of \cite{deWit} the pullback of the three-form components are,
\begin{eqnarray}
C^{(10)}_M &=& \frac{1}{2}\epsilon^{uv}\partial_u X^N\partial_vX^LC^{(10)}_{MNL} -\epsilon^{uv}\partial_u X^-\partial_vX^NC^{(10)}_{-MN}, \\
C^{(10)}_{\pm}&=& \frac{1}{2}\epsilon^{uv}\partial_uX^M\partial_vX^N C^{(10)}_{\pm MN} , \label{Cmasmenos} \\
C^{(10)}_{+-} &=& \epsilon^{uv}\partial_uX^-\partial_v  X^N C^{(10)}_{+-N}.
\end{eqnarray}
By performing a Legendre transformation we obtain the canonical Hamiltonian
\begin{eqnarray}
\mathcal{H}_{CAN} &=& \frac{1}{2(P_--C^{(10)}_- -B_{-})}\left[(P_M-C^{(10)}_M-B_M)^2 + \Pi^u\Pi^v\gamma_{uv} + {G}\right], \nonumber \\
&-& B_+  +\Pi^u\partial_u A_0-C^{(10)}_{+-}-C^{(10)}_+,
\end{eqnarray}
subject to the primary constraints
\begin{eqnarray}
\varphi &=& \Pi^0 \approx 0 , \\
\phi_{u} &=&  P_M\partial_uX^M + P_-\partial_uX^-+\Pi^vF_{uv} \approx 0,
\end{eqnarray}
with 
\begin{eqnarray}
B_M&=&\Pi^u(\partial_u X^-B_{M-}+\partial_uX^{N}B_{MN}), \label{BM} \\
B_+ &=& \Pi^u(\partial_u X^-B_{+-}+\partial_uX^N B_{+N}). \label{Bmasmenos}
\end{eqnarray}
 From the preservation in time of $\varphi$ the Gauss constraint is obtained (\ref{Gaussconstraint}) as a time independent secondary constraint. Moreover, $\phi_u$ is also time independent and the three constraints are first class. 
 
By fixing the residual gauge symmetries related to the RR three-form transformation as well as the two form NSNS, it can be set up $B_{+-}=B_{-M}=C^{(10)}_{+-M}=0$.  $C^{(10)}_{-MN}=0$ can be fixed to zero. If we choose a background where $C_{-MN}$ is not zero but constant, once a quantization condition is imposed on $C^{(10)}_{-}$, it is no longer possible to take continuously this value to zero. 
Analogously to the methodology discussed in section (\ref{Sec3}), it is possible to find a suitable canonical transformation in which the D2-brane Hamiltonian may be re-expressed in terms of the physical variables, 
\begin{eqnarray}
\widehat{P}_M&=& P_M-C^{(10)}_M , \\
\widehat{P}_- &=& P_--C^{(10)}_-.
\end{eqnarray}
This is a consistent canonical transformation that preserves all the Poisson brackets of the theory and the kinematical contribution. 
In order to obtain the physical Hamiltonian, the gauge invariance of the theory has been fixed according to
\begin{eqnarray}
& &\Pi^0 \approx 0 \rightarrow \psi_1=A_0\approx 0 , \\
& & \phi_\omega \approx 0 \rightarrow \psi_2= \Sigma_u = \gamma_{0u}+(\partial_0A_v + B_{0v} )\gamma^{vw}\mathcal{F}_{uw} \approx 0.
\end{eqnarray}
While the first condition leaves no residual symmetry the second one leaves the expected APD constraint. Let us remark that we have left unfixed the gauge symmetry related with the Gauss constraint. Finally, the LCG D2-brane DBI physical Hamiltonian coupled to RR and NSNS background fields is given by
\begin{eqnarray}\label{H DBI RR NSNS}
\hspace{-0.5cm}H&=& \int d^2\sigma \left[ \frac{1}{2}\frac{(\widehat{P}_M - B_M)^2}{\sqrt{W}} + \frac{1}{2}\frac{\Pi^u\Pi^v\gamma_{uv}}{\sqrt{W}}+\frac{1}{2}\frac{{G}}{\sqrt{W}}-C^{(10)}_+ - B_{+}\right],
\end{eqnarray}
subject to the Gauss Law and the APD constraint respectively
\begin{eqnarray}
\partial_u\Pi^u &\approx& 0, \label{Gauss Constraint}\\
\epsilon^{uv}\partial_u\left[ \frac{P_M}{\sqrt{W}}\partial_vX^M + \frac{\Pi^w}{\sqrt{W}}F_{vw} \right] &\approx& 0 \label{APD Constraint},
\end{eqnarray}
where $C^{(10)}_{+}$ is given by (\ref{Cmasmenos}) and $B_M$, $B_{+}$ are given by (\ref{BM}), (\ref{Bmasmenos}), respectively (with $B_{-M}=B_{+-}=0$). Therefore, the two resultant symmetries of the theory are the expected Gauss constraint (\ref{Gauss Constraint}), associated to the gauge U(1) and the Area Preserving Diffeomorphisms symmetries (\ref{APD Constraint}), both related to the worldvolume of the D2-brane. 
By using the equations of motion it can be seen that the gauge $\Sigma_u = 0$ fixes the Lagrange multiplier $c^w=0$ and, in consequence $\widehat{P}_-$ can be written as a scalar density 
 \begin{eqnarray}
 \widehat{P}_-=\widehat{P}^0_-\sqrt{W},
 \end{eqnarray}
 where, as in the case of the nontrivial M2-brane, $P_-^0$ is a constant and $\sqrt{W}$ is the regular density on $\widetilde{\Sigma}$. It is straightforward to see that turning off all the background fields, one recovers the LCG Hamiltonian of the D2-brane studied in \cite{Manvelyan1}.

In order to reproduce the LCG Hamiltonian (\ref{HCMIM22}) found in Section (\ref{Sec3}), we consider a toroidal compactification of (\ref{H DBI RR NSNS}) and impose particular independent quantization conditions. The transverse index $M=(\alpha,r)$ can be decomposed in $\alpha = 3,\dots,8$ non compact directions and $r,s=1,2$ compact directions. The topology of the compact D2-brane worldvolume from now on, will be assumed to be also a 2-torus. 
In analogy with the M2-brane, the winding condition on the map on the compact sector is 
\begin{eqnarray}\label{M-tilde}
\oint_{\widetilde{C}_{s}} dX = 2\pi \widetilde{R}(l^{'}_s+m^{'}_s\widetilde{\tau})=\widetilde{M}_s^1 + i \widetilde{M}_s^2
\end{eqnarray}
where $\widetilde{R},\widetilde{\tau}$ are moduli of $T^2$ and $l'_s$, $m'_s$ with $s=1,2$ are winding numbers. Therefore, the harmonic sector of the map can be written in terms of a normalized basis $d\widehat{X}^s$ as $dX_h=dX_h^1 + idX_h^2= (\widetilde{M}_s^1 + i\widetilde{M}^2_s) d\widehat{X}^s$.

Since $\mathcal{F}_{uv}^{DBI}$ contains the pullback of the NSNS background field $B$,  $\mathcal{F}$  becomes, \color{black}
\begin{eqnarray}\label{FDBI}
\mathcal{F}^{DBI}=\frac{1}{2}\epsilon^{u\bar{u}}\epsilon^{v\bar{v}}\mathcal{F}_{\bar{v}\bar{u}}^{U(1)}\mathcal{F}_{vu}^{DBI} = \widetilde{\mathcal{F}}^{DBI}+ \epsilon^{\bar{u}u}\epsilon^{\bar{v}v} F_{\bar{v}\bar{u}}A_{vu} +\epsilon^{\bar{u}u}\epsilon^{\bar{v}v} A_{\bar{v}\bar{u}vu},
\end{eqnarray}
with
\begin{eqnarray}
A_{vu} &=& 2\partial_v X^\alpha \partial_u X^r B_{\alpha r} + \partial_v X^r \partial_u X^s B_{rs} , \label{Acurvatura} \\
A_{\bar{v}\bar{u}vu}&=& 2\partial_{\bar{v}} X^\alpha \partial_{\bar{u}} X^\beta B_{\alpha\beta} \partial_{v} X^{\bar{\alpha}} \partial_{u} X^r B_{\bar{\alpha}r} + \partial_{\bar{v}} X^\alpha \partial_{\bar{u}} X^\beta B_{\alpha\beta} \partial_v X^r \partial_u X^s B_{rs}, \nonumber \\
&+&2\partial_{\bar{v}} X^\alpha \partial_{\bar{u}} X^r B_{\alpha r} \partial_v X^{\bar{\alpha}} \partial_u X^r B_{\bar{\alpha}r} + 2\partial_{\bar{v}} X^\alpha \partial_{\bar{u}} X^r B_{\alpha r} \partial_v X^r \partial_u X^r B_{rs} , \nonumber \\
&+& \partial_{\bar{v}} X^r \partial_{\bar{u}} X^s B_{rs} \partial_v X^{\bar{r}} \partial_u X^{\bar{s}} B_{\bar{r}\bar{s}}, \label{Acurvatura2}
\end{eqnarray}
and  $\widetilde{\mathcal{F}}^{U(1)}$ represents the determinant of the $U(1)$ DBI curvature associated to the contribution related to the non-compact sector. 

We find that the LCG Hamiltonian on $M_8\times T^2$ is given by
\begin{eqnarray}\label{compactD22}
H&=& \int d^2\sigma \left[ \frac{1}{2}\frac{(\widehat{P}_\alpha - B_\alpha)^2}{\sqrt{W}} +\frac{1}{2}\frac{(\widehat{P}_r - B_r)^2}{\sqrt{W}}+ \frac{1}{2}\frac{\Pi^u\Pi^v \gamma_{uv}}{\sqrt{W}}+\frac{1}{2}\frac{\widetilde{G}}{\sqrt{W}} \right., \nonumber \\
&+& \left. \frac{1}{2}\frac{\epsilon^{u\bar{u}}\epsilon^{v\bar{v}}F_{\bar{v}\bar{u}}A_{vu} }{\sqrt{W}} + \frac{1}{2}\frac{\epsilon^{u\bar{u}}\epsilon^{v\bar{v}}A_{\bar{v}\bar{u} vu} }{\sqrt{W}} + \frac{1}{2}\sqrt{W}\left\lbrace X^\alpha,X^r\right\rbrace^2 +\frac{1}{4}\sqrt{W}\left\lbrace X^r,X^s\right\rbrace^2, \right.   \nonumber \\
&-&\left. \frac{1}{2}\epsilon^{uv}\partial_uX^r\partial_v X^s C^{(10)}_{+rs} - \Pi^u \partial_u X^\alpha B_{+\alpha} - \Pi^u \partial_u X^r B_{+r}   \right].
\end{eqnarray}
subject to
\begin{eqnarray}
\partial_u\Pi^u &\approx& 0, \\
\epsilon^{uv}\partial_u\left[ \frac{P_\alpha \partial_v X^\alpha}{\sqrt{W}} + \frac{P_r \partial_vX^r}{\sqrt{W}} + \frac{\Pi^w F_{vw}}{\sqrt{W}} \right] &\approx& 0,\\
 \oint_{C_S}\left[\frac{P_\alpha \partial_v X^\alpha}{\sqrt{W}} + \frac{P_r \partial_vX^r}{\sqrt{W}} + \frac{\Pi^w F_{vw}}{\sqrt{W}}\right] d^v\sigma &\approx& 0,
\end{eqnarray}
where ${\widetilde{G}}=det(\widetilde{\gamma}_{uv} + \widetilde{\mathcal{F}}_{uv}^{U(1)})$ represents the generalized metric associated to the noncompact dimensions and with 
\begin{eqnarray}
\widetilde{\mathcal{F}}_{uv}^{U(1)} &=& F_{uv}  +\partial_u X^\alpha \partial_v X^\beta B_{\alpha\beta}, \\
\widetilde{\gamma}_{uv} &=&\partial_u X^\alpha \partial_v X_\alpha.
\end{eqnarray}
In order to make contact with (\ref{HCMIM22}) it is required to assume the RR three-form background to be constant  $C_{\pm rs}^{(10)}=\epsilon_{rs}c_{\pm}$ an quantized. It implies the existence of a well-defined closed two form $\displaystyle \widetilde{F}_{\pm}=\frac{1}{2}C_{\pm rs}^{(10)}\widetilde{M}^r_p\widetilde{M}^s_q d\widetilde{X}^p\wedge d\widetilde{X}^q$  with $\widetilde{M}^r_p$ defined as in (\ref{M-tilde}) and with $\widetilde{X}^p$, $p=1,2$, the $T^2$ coordinates.  If a quantization condition on $T^2$ is imposed, this implies the presence of a nontrivial 2-form flux on the D2-brane worldvolume,
\begin{eqnarray}
\label{FluxT22}
\int_{T^2} \widetilde{F}_{\pm}= k_\pm\in \mathbb{Z}\ne 0 \to c_{\pm}\int_{\widetilde{\Sigma}}  \widehat{F} = k_\pm\ne 0
\end{eqnarray}
where $k_{\pm}=nc_{\pm}$ and $\displaystyle \widehat{F}=\frac{1}{2}\epsilon_{rs}dX_h^r\wedge dX_h^s$. We can also impose a quantization condition the pullback of the NS-NS background two-form $\widetilde{B}_2$ on $T^2$. Since the B-field is also assumed constant its pullback  can also be interpreted as a flux condition over the worldvolume,
\begin{eqnarray}\label{flux B}
 \int_{T^2}\widetilde{B}_2=k_B\ne 0 \to \int_{\widetilde{\Sigma}} B_2 = b_2\int_{\widetilde{\Sigma}} \widehat{F}= k_B\ne 0,\quad k_B \in \mathbb{Z},
\end{eqnarray}
 with $\widetilde{B}_2=\frac{1}{2}B_{rs} \widetilde{M}^r_p \widetilde{M}^s_q d\widetilde{X}^p\wedge d\widetilde{X}^q$, $B_{rs}=b_2\epsilon_{rs}$ and $k_B=nb_2$. Both quantization conditions (\ref{FluxT22}) and (\ref{flux B}) imply the existence of an irreducible wrapping condition on the compact sector.
 In this context, the imposition of the quantization conditions over the $B_2$  and $C^{(10)}_{\pm}$ are completely independent. However, from the CM2-brane dual description,  we have shown that they they are associated to different components of the 11d three-form $C_3$, hence both of them must hold in order to describe its  M2-brane dual origin. 

The background can be fixed such that the only nontrivial components of the constant background fields be $B_{rs}, C^{(10)}_{\pm rs}, B_{+r}$. Then, it can be seen from (\ref{Acurvatura}) and (\ref{Acurvatura2}) that
\begin{eqnarray}
\epsilon^{\bar{u}u}\epsilon^{\bar{v}v} F_{\bar{v}\bar{u}}A_{vu} &=& (\sqrt{W})^2 (\star F)\left[\widehat{F}^{rs} + \mathcal{F}^{rs}\right]B_{rs}, \nonumber \\
\epsilon^{\bar{u}u}\epsilon^{\bar{v}v} A_{\bar{v}\bar{u}vu} &=&
(\sqrt{W})^2\frac{1}{2}\left(\widehat{F}^{rs}B_{rs} + \mathcal{F}^{rs}B_{rs}\right)^2.
\end{eqnarray}
with $F=dA$. Finally, if the moduli,  winding numbers and units of fluxes are those considered in Section 4, the LCG Hamiltonian of a D2-brane on $M_8\times T^2$ subject to the 2-form fluxes conditions (\ref{FluxT22}) and (\ref{flux B}) exactly becomes 
\begin{eqnarray}\label{HCMIM222}
\mathcal{H}_{dual} &=& \frac{1}{2} \frac{\widehat{P}_\alpha  \widehat{P}^\alpha}{\sqrt{W}} + \frac{1}{2}\frac{(\widehat{P}_r - B_r)^2}{\sqrt{W}} + \frac{1}{2}\frac{\Pi^u\Pi^v \widetilde{\gamma}_{uv}}{\sqrt{W}} + \frac{G}{2\sqrt{W}} + \sqrt{W}\frac{1}{4}(\widehat{F}^{rs})^2  \nonumber \\
&+& \frac{1}{4}\sqrt{W}(\mathcal{F}^{rs})^2+ \frac{1}{2}\sqrt{W}(\mathcal{D}_r X^\alpha)^2 - C_{+}^{(10)} - B_+
\end{eqnarray}
where $G$, $\mathcal{F}^{rs}$ and $\mathcal{D}_r$ coincides with the definitions used in (\ref{HCMIM22}). The Hamiltonian is also subject to the Gauss constraint
(\ref{Gaussconstraint}) and to worldvolume symplectomorphisms (\ref{APD1}) and (\ref{APD2}). 

There are several differences between this formulation and a general D2-brane with RR and NSNS fluxes: First,  the background fields are considered constant and when the quantization condition is imposed. They are responsible for generating specific 2-form fluxes whereas this statement is not necessarily true in more general backgrounds. These fluxes imply the existence of a monopole contribution over the D2-brane worldvolume in analogy with \cite{Restuccia} that acts as an extra constraint on the D2-brane embedding maps in distinction with any other case considered so far. It implies, on top of the flux contribution, the presence of an extra symplectic gauge field that couples to the scalar fields via a symplectic covariant derivative defined $\mathcal{D}_r X^\alpha$ and a symplectic curvature $\mathcal{F}_{rs}$. There are also cubic and quadratic interactions between the different field strengths here defined, which are not present in any of the previous cases studied in the literature, up to our knowledge. 
%In particular, the compactified momentum has a nontrivial coupling to the worldvolume B-field. 
As a result, the  Hamiltonian (\ref{HCMIM22}) obtained through  a scalar/vector dualization of the CM2-brane, corresponds to a specific D2-brane with constant quantized background fields that induce 2-form flux acting on the worldvolume as an extra constraint which generate new fields and couplings in the Hamiltonian. 

%%%%%%%%%%%%%%%%%%%%%%%%%%%%%%%%%%%%%%%%%%%%%%%%%%%%%%%%%%%%%%%%%%%%%%%%%%%%%%%%%%%%%%%%%%%%%%%%%%%%%%%%%%%%%%%%%%%%

\section{Discussion and Conclusions}
We have obtained the LCG Hamiltonian from a toroidally compactified M2-brane with 2-form $C_{\pm}$ fluxes, with an explicit contribution of the transverse components of supergravity three-form, $C_{abc}$ denoted by us as CM2-brane. The direct analysis of its spectral properties is rather cumbersome due to the mixing terms between the scalar fields with their canonical conjugate momenta that appear in the kinetic term. In fact, the sufficiency criteria for discreteness of the supersymmetric spectra found in \cite{Boulton} is not directly applicable.  However, we obtain a canonical transformation of the phase space variables that establishes an equivalence with  the M2-brane with $C_{\pm}$ fluxes formerly identified in \cite{mpgm6}. Hence, they must share the same spectral discreteness properties and consequently, the CM2-brane also constitutes a nontrivial M2-brane. The M2-brane with central charge  is obtained by performing the canonical transformation (\ref{ct3}) and turning off the $C_+$ %and $C_{abc}$
components. The toroidally nontrivial M2-branes here analyzed are shown to contain the same type of $U(1)$ monopole quantization condition. 

We now obtain the D-brane description of the CM2-brane Hamiltonian. We find that it corresponds to a D2-brane in the presence of two-form fluxes associated to the quantization of constant RR and NSNS background fields. They appear in the dualization of the quantization condition of the 11D $C_{\pm}$. Due to the new terms in the CM2-brane formulation, the DBI B-field contribution becomes explicit in the D-brane counterpart. The compactified momentum has a nontrivial coupling to the worldvolume B-field. By using the relation (\ref{quantization}), the $C_{\pm}^{(10)}$ flux condition and the NSNS quantized $B_+$ are straightforwardly obtained. Gauge invariance allows to set $B_-=0$. The NSNS constant $B_2$ field becomes quantized without imposing an extra quantization condition since it is proportional to the central charge condition. This dual\footnote{By means of a scalar/vector dualization.} D-brane theory inherits from the nontrivial M2-brane, the same type of $U(1)$ monopole quantization condition. Since it acts as an extra constraint on the  Hamiltonian, it also implies the appearance of a dynamical symplectic gauge field associated to $d\mathcal{A}^r$ components with a topologically trivial curvature in analogy with the M2-brane with central charges. The D-brane theory is  subject to the Gauss and APD constraints. The theory contains new symmetries. It is possible to define an extra $U(1)$ gauge symmetry under symplectomorphism transformations in terms of the components of the embedding maps $\mathcal{A}_G$. $\mathcal{A}$ and $\mathcal{A}_G$ define respectively a symplectic curvature $\mathcal{F}$ and a $U(1)$ curvature $\mathcal{F}_G$. As happens with the CM2-brane, when expressed in terms of the embedding map components, they have the unique property of being equal.

 The $\mathcal{A}_G$ together with the topologically nontrivial $U(1)$ gauge symmetry associated with the fluxes, $\widehat{A}$,  defines a dynamical $U(1)$ gauge symmetry $\mathbb{A}$ on a nontrivial $U(1)$ fiber bundle with the same Chern class as the flux condition. These two fibers are linked together to form a twisted torus. The D2-brane bundle is defined by a twisted torus bundle with monodromy in $SL(2,Z)$ -inherited from the toroidally nontrivial M2-brane- and an independent  DBI $U(1)$ trivial principal bundle over  its worldvolume. 

The quantization condition of a constant $\widetilde{B}_2$ was formerly discussed \cite{Connes} in the context of the noncommutative formulation of the matrix model on a torus and its M-theory origin was qualitatively discussed in terms of the coupling of the supermembrane to a constant $C_{-}$ in \cite{deWit}. The supermembrane on a $C_-$ background was already discussed in \cite{mpgm6}. The $B_2$ contribution does not come from the dualization of the $C_-$ but from the $C_{abc}$. Hence the non commutative formulation of matrix model on a torus has an M-theory origin associated to a CM2 for a restricted background where $B_+$ and $C_+$ vanishes.  Both formulations  can generate a monopole contributions. 

Because the D2-brane carries 2-form fluxes, an open question is whether it admits a description in terms of Dp bound states. The D2-D0 bound state would have been a natural candidate. In \cite{mpgm11}  the spectral properties of the $SU(N)$ regularized M2-brane with central charge and the $0+1$ dimensionally reduced D2-D0 bound state were compared.  Though both models seem quite similar and carry a RR charge, they are associated to different monopole conditions and their spectra are completely different. Furthermore in the CM2 dual (\ref{HCMIM22}) the DBI curvature $F$ does not get quantized in distinction with the D2-D0 bound state. Hence, the M2-brane dual considered in this work can not be described as a D2-D0 bound state. This analysis however does not exclude the possibility of more complicated constructions, as for example in \cite{mpgm20}. A deeper analysis of this aspect is left for a future work. 

\section{Acknowledgements}
MPGM and CLH thanks to A. Restuccia for helpful discussions. CLH is supported by CONICYT PFCHA/DOCTORADO BECAS CHILE/2019-21190263, and the Project ANT1956 of the U.Antofagasta. The authors also thank SEM 18-02 funding project from U. Antofagasta, and to the international ICTP Network NT08 for kind support.

% \bibliography{references}

\end{document}